\PassOptionsToPackage{draft}{hyperref}
\documentclass[runningheads]{llncs}
\usepackage{todonotes}
\usepackage{graphicx}
\usepackage{hyperref}
\begin{document}

\newcommand{\repeatthanks}{\textsuperscript{\thefootnote}}
\title{VR Accessibility in Distance Adult Education}

%
%
\author{Bartosz Muczyński\thanks{The first two authors contributed equally to this study.}\inst{2}\orcidID{0000-0002-0559-4181} \and
Kinga Skorupska\repeatthanks \inst{1}\orcidID{0000-0002-9005-0348} \and
Katarzyna Abramczuk\inst{5}\orcidID{0000-0003-2249-9888}  \and
Cezary Biele\inst{3}\orcidID{0000-0003-4658-5510} \and
Zbigniew Bohdanowicz\inst{3}\orcidID{0000-0002-5430-0485}  \and
Daniel Cnotkowski\inst{3}\orcidID{0000-0002-9009-8018}  \and
Jazmin Collins \inst{6}\orcidID{0000-0001-6850-5820} \and
Wiesław Kopeć \inst{1}\orcidID{0000-0001-9132-4171} \and
Jarosław Kowalski\inst{3}\orcidID{0000-0002-1127-2832}  \and
Grzegorz Pochwatko\inst{4}\orcidID{0000-0001-8548-6916} \and
Thomas Logan\inst{7} }

\authorrunning{Muczyński and Skorupska et al.}
%

\institute{XR Lab, Polish-Japanese Academy of Information Technology (PJAIT)
\email{kinga.skorupska@pja.edu.pl}\\
\and
Faculty of Navigation, Maritime University of Szczecin
\email{b.muczynski@pm.szczecin.pl}\\
\and
Laboratory of Interactive Technologies, National Information Processing Institute (LIT, OPI PIB)
\and
VR Lab, Institute of Psychology Polish Academy of Sciences (IP PAS)
\and
University of Warsaw, Poland
\and
Enhancing Ability Lab, Cornell University
\and
Equal Entry LLC.}
\maketitle              
\begin{abstract}
As virtual reality (VR) technology becomes more pervasive, it continues to find multiple new uses beyond research laboratories. One of them is distance adult education---the potential of VR to provide valuable education experiences is massive, despite the current barriers to its widespread application. Nevertheless, recent trends demonstrate clearly that VR is on the rise in education settings, and VR-only courses are becoming more popular across the globe. This trend will continue as more affordable VR solutions are released commercially, increasing the number of education institutions that benefit from the technology. No accessibility guidelines exist at present that are created specifically for the design, development, and use of VR hardware and software in distance education. The purpose of this workshop is to address this niche. It gathers researchers and practitioners who are interested in education and intend to work together to formulate a set of practical guidelines for the use of VR in distance adult education to make it accessible to a wider range of people.

\end{abstract}
\section{Overview}
\subsection{Theme}

Barriers to learning in distance education are long established, from the onset of the correspondence courses of the nineteenth century, through education via the radio in the twentieth century, to 
VR-mediated classrooms \cite{CLARK2020410} that focus on teaching practical skills in the twenty-first century \cite{VrDistanceEducation_2022}. These barriers include, most notably, the absence of institutional support such as staff training, inadequate technological preparation and infrastructure, inadequate policies and negative stereotypes \cite{Yeh2022-br}. A recent large-scale exploratory study found ``(1) administrative structure, (2) organizational change, (3) technical expertise, (4) social interaction and quality, (5) faculty compensation and time, (6) threat of technology, (7) legal issues, (8) evaluation/effectiveness, (9) access, and (10) student‐support services'' to be the ten key barriers to distance education \cite{linfactorsbarriers_2021}. Although the technology has developed in terms of extra-course factors, such as access, network stability, technological readiness, and the support policies of educational institutions, accessibility issues continue to hinder both virtual reality (VR) software functionality and VR instructional design. 
Many VR social applications lack accessibility features and support for educational settings, despite their use in education and the existence of VR-specific accessibility guidelines\footnote{This includes the guidelines of W3C, available here: \url{https://www.w3.org/TR/xaur/}, guidelines for developers by Oculus \url{https://developer.oculus.com/resources/design-accessible-vr/}, and considerations related to professional settings \url{https://xraccess.org/bcxr-report/}}. To address this niche during this workshop, we aspire to gather and analyze diverse perspectives, experiences, and approaches to VR distance education in the context of accessibility. We hope that the resulting practical guidelines for making VR more accessible in distance education will help researchers, practitioners, and educators to better design their VR-mediated distance learning experiences, and will provide a valuable perspective for software developers that will help them prioritize functional features that are crucial to the solutions' accessibility in VR distance education. This, hopefully, will serve as a step toward the application of Universal Design for Learning for new and for existing XR educational environments; that, in turn, will make those environments better not only for users with disabilities, but for all users (via so called curb-cut effect \cite{hessecurbcut1995}).  
It is worth noting that in 2020 the Game Awards, which is the most prestigious and recognizable event in the video game industry, introduced the Innovation in Accessibility award to recognize achievements made by the largest game studios in making their games more accessible. This further demonstrates that VR education applications lag behind the entertainment industry in their support for diverse groups of users.   

\subsection{Target audience}
We invite researchers and practitioners who represent the fields of distance education, virtual reality, and accessibility, as well as those with combined experience in these areas. We encourage people that are interested in new forms of distance education to join our debate on the opportunities and challenges of the application of VR in their fields---even if they lack prior experience with the technology. 

\subsection{Contributions}

Participants will prepare short abstracts (2--5 pages) containing:\begin{itemize}
    \item a short bio that includes the participant's experience in distance education, virtual reality, and accessibility
    \item at least one of the following topics:
    \begin{itemize}
\item Real-life examples of in-person or software-mediated educational situations in which specific accessibility features were absent or sorely needed.
\item The pros and cons of specific tools and applications (e.g. text, video, or VR) for distance education that the participant has used before. This point should be as detailed as possible with special attention paid to accessibility features.
\item Areas in which VR-based distance education can be more accessible than standard (in-person) or remote video education. 
\item A description of powerful examples of VR accessibility features that the participant has used or encountered in the broad range of VR applications, including VR games and experiences.
\item An idea for a single new or modified software or hardware feature that could have a positive impact in the creation of an accessible VR-based distance course or lesson.
\item An idea for a course or lesson that the participant would like to conduct in VR, including the goal of the course or lesson, a detailed schedule (topics and number of hours), teaching methods and activities, forms of evaluation (e.g. assignments or tests/quizzes), and a reflection of challenges the participant might encounter while conducting it.
    \end{itemize}
\end{itemize}
Participants may also consider situations in which, temporarily or permanently, not all attendees have access to VR headsets, which means that they must join using alternative means, such as voice, text, or video communication using a desktop or a mobile phone with limited data.



\subsection{Topics}
In this workshop, we are interested in the interplay of three key aspects:
\begin{enumerate}

\item Software solutions. Participants will consider a selection of major applications and platforms, such as Horizon Workrooms and Engage, as well as the influence of aspects such as environments, avatars, and available interactions in the VR experience\cite{HARFOUCHE20201187}. This part will enable us to establish a set of features that are necessary for accessible  distance education in the VR environment, optional features that impact the education experience positively, and features that could be implemented to enhance or extend education experiences. 
\item Hardware and its capabilities. During the workshop, participants will identify the strengths and weaknesses of each major commercially available solution in the context of accessibility for distance education. Based on the identified weaknesses, participants will work to develop countermeasures that could be implemented in the next generation of VR hardware. 
\item Instructional design for VR. Participants will evaluate how the design of courses for VR classrooms ought to differ from traditional learning, and from asynchronous and synchronous e-learning. This topic will be based on the findings of the sections above and will focus on interaction design guidelines. 
\end{enumerate}

\section{Organisers}

\hspace{\parindent} Katarzyna Abramczuk works at the Faculty of Sociology of the University of Warsaw. She is interested in research at the intersection of sociology, mathematical modeling of social phenomena, cognitive psychology, behavioural economics, and new technologies. 

Cezary Biele is the head of the Laboratory of Interactive Technologies at OPI PIB. He is a human--technology interaction researcher with a background in psychology, psychophysiology, and computer science. 

Zbigniew Bohdanowicz is a researcher at the Laboratory of Interactive Technologies at OPI PIB. He works on the social aspects of technology development---specifically, its impact on individuals and social networks. 

Daniel Cnotkowski is a VR programmer at the Laboratory of Interactive Technologies at OPI PIB. He works on VR applications for researchers. He is interested in how VR can become a mainstream platform for research simulations. 

Jazmin Collins is a PhD student at the Enhancing Ability Lab of Cornell University. Her research focuses on both the evaluation and the development of emerging VR/augmented reality (AR) technologies as accessible tools for blind/low vision communities. She is also a member of the \href{https://xraccess.org}{XR Access} organization, which promotes accessibility in VR/AR.

Wiesław Kopeć is a computer scientist and the head of XR Center at PJAIT and a member of Emotion-Cognition Lab at SWPS University. He also co-founded the transdisciplinary Human Aspects in Science and Engineering (HASE) research group and distributed LivingLab Kobo.

Jarosław Kowalski is an assistant professor at the Laboratory of Interactive Technologies at OPI PIB. He is a sociologist and research specialist. His areas of interest include User Experience studies, the sociology of innovation, and the social and psychological aspects of new technologies. 

Thomas Logan is the owner of Equal Entry and has spent the past twenty years assisting organizations to create technology solutions that work for people with disabilities. He has delivered projects for numerous federal, state, and local government agencies as well as private sector organizations.

Bartosz Muczyński is a researcher at the Maritime University of Szczecin, Poland, where he leads the laboratory of VR and AR systems and serves as the head of the university's E-learning Center. He also works as a VR developer at OPI PIB.

Grzegorz Pochwatko is a leading expert in VR and psychophysiology. As the head of the Virtual Reality and Psychophysiology Lab at IP PAS, he has dedicated his career to exploring the interactions between virtual humans and co-presence in extended reality environments. 

Kinga Skorupska is a researcher at the XR Center at PJAIT whose interests include UX Design and science communication. She conducts research in collaborative ICT solutions and technology enhanced learning.



\section{Methodology}
We propose a one-day remote hybrid VR and video workshop. The workshop will commence with a Zoom video call for introductions and ice-breaking. The participants will then share insights from their submitted abstracts, which contain specific cases and reflections they had on the subject. A discussion of recurring insights will follow. Participants will also be prepared for the next session, which will take place in a VR education environment. The organizers will explain how the participants should approach it, what to do in the case of any technical issue, and what the main goals of the experience are.

The second session, in Horizon Workrooms, will focus on the delivery of a hybrid VR/video experience to the participants. Although participants are encouraged to join using head-mounted displays, it will also be possible to use a desktop application to join the VR experience via video call. Having a part of the group joining the VR space via 2D video call will help expand the experience and perspectives of people that do not have access to VR headsets. Participants will be given a brief lecture on the W3C XR Accessibility guidelines, which will serve as an introduction to a group task in which each group will work with a few selected guidelines based on personas constructed previously from case studies gathered by the organizers. This will be complemented with a discussion and brainstorming session for all participants, during which all ideas will be compiled on a board.  

In the next session, the participants will reflect on their shared VR education experience, and a facilitator will help them brainstorm insights that will later be categorized using affinity diagramming. The third session will focus on distance education functionalities, such as moderation for teachers, student evaluation, and note taking. Participants will discuss how the ideal education VR software should be designed to provide teachers with a set of features that support the teaching process and give due consideration to the accessibility features identified during the first session. The workshop will conclude with a discussion on a possible multi-author publication, a follow-up action plan, and a schedule for related activities. 



\begin{figure}
\includegraphics[width=\textwidth]{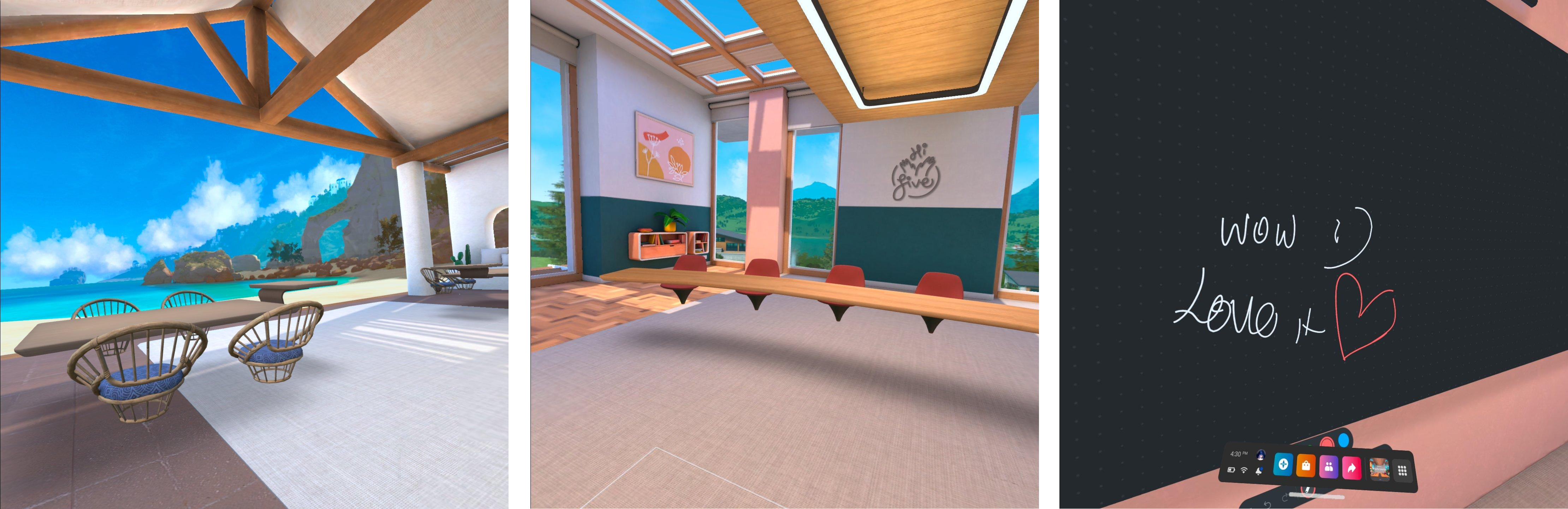}
\caption{We have selected Horizon Workrooms as our virtual venue for Sessions 2 and 3 of the Workshop to enable participants to experience distance education via VR, or, if impossible, by joining a predominantly VR session via a desktop application.} \label{virtualvenue}
\end{figure}

\section{Expected Outcomes}
The key outcome of the workshop will constitute a set of practical accessibility guidelines for the use of VR in distance education, which will assist developers, researchers, practitioners, and educators in better designing their VR-mediated distance learning experiences and applications. Although we expect these guidelines to be inspired by existing accessibility guidelines in VR and beyond, they may also draw heavily from both the diverse knowledge and experience of the participants, and relevant lessons highlighted in prior literature. Such guidelines will be particularly useful for practitioners of distance education who lack prior VR experience or expertise. The guidelines will be compiled in a post-workshop multi-author publication coordinated by the workshop organizers. The publication will be submitted to an ACM venue to ensure high visibility to other researchers. This justifies another connected outcome: the dissemination of accessibility concerns in connection to XR technology that is not limited to instructional settings. Both outcomes connect directly to the theme of this years conference, ``Design for Equality and Justice".

\bibliographystyle{splncs04}
\bibliography{mybibliography}

\clearpage
\section*{Workshop Schedule and Activities (Appended Post-publication by Authors)} 

\footnotesize
\textbf{Requirements:} Although we encourage participants to use Oculus Quest 1, Meta Quest 2 (Formely Oculus Quest 2), or Meta Quest Pro to join the Horizon Workrooms Session via VR, it is also possible for them to join solely via video using a computer.\\
\textbf{Pre-workshop: } A test session in Horizon Workrooms in VR will be held two days prior to the workshop to allow participants to test their setups and to become familiar with the environment before joining the VR session on the day of the workshop. \\
\noindent
\textbf{Participants: }12--20 (max. 16 via VR)


\subsection{Part 1: Challenges and opportunities in using VR for diverse education cases (2 hours)} 
\textbf{Goal:} Explore contexts for VR distance education. \\
\\
\textbf{Activities during a Zoom video call:}
\begin{itemize}
\item Ice-breaking: Welcome, introductions, and the plan of the workshop (15 minutes)
\item Brief presentations of selected abstracts by participants (75 minutes)
\item Moderated discussion and affinity diagramming of key insights (30 minutes)
\end{itemize}

\subsection*{Coffee break (15 minutes) }

\subsection{Part 2: Simulation of learning experience elements in hybrid VR/video mode (1 hour)}
\textbf{Goal:} To provide participants with a shared diverse VR learning experience using the most common education activities \\
\\
\textbf{Activities in VR/video via Horizon Workrooms:}
\begin{itemize}

\item Lecture: An overview of VR hardware and software accessibility guidelines (20 minutes)
\item Work in groups: Analysing the accessibility challenges and opportunities of VR software based on case-study-informed personas (20 minutes)
\item Brainstorming: Gathering group insights on the shared board (20 minutes)
\end{itemize}

\subsection*{Lunch break (60 minutes)}

\subsection{Part 3: Conclusions and Follow-up (2 hours)}
\textbf{Goal:} To formulate guidelines for VR distance education and to discuss the next steps. \\
\\
\textbf{Activities in a Zoom video call:}
\begin{itemize}
\item Group work: brainstorming potential software features based on challenges faced by the user personas (30 minutes)
\item Presentations of group work outputs with discussion (70 minutes)
\item Discussion of the next steps for the post-workshop multi-author publication (10 minutes)
\item Final remarks and thank-yous (10 minutes) \\
\end{itemize}

\end{document}